\documentclass[12pt]{article}
\usepackage{graphicx}

\def\pbnr{}
\def\speaker{Mohammad Alhakami}
\def\onbehalfof{}
\def\title{Odd- and even-parity charmed mesons revisited in heavy hadron chiral perturbation theory}
\def\affiliation{School of Physics and Astronomy\\
The University of Manchester, Manchester, UK}
\def\support{The workshop was supported by the University of Manchester, IPPP, STFC, and IOP}

\newcommand\pubnumber{\pbnr}
\newcommand\pubdate{\today}
\def\Title#1{\begin{center} {\Large #1 } \end{center}}
\def\Author#1{\begin{center}{ \sc #1} \end{center}}

\newcommand{\OnBehalf}[1]{\sbox0{#1}\ifdim\wd0=0pt
        {}
	\else
	{\\on behalf of #1}
	\fi}
\newcommand{\SupportedBy}[1]{\sbox0{#1}\ifdim\wd0=0pt
        {}
	\else
	{\footnote{#1}}
	\fi}
\def\Address#1{\begin{center}{ \it #1} \end{center}}

\newcommand\pubblock{\includegraphics[width=5cm]{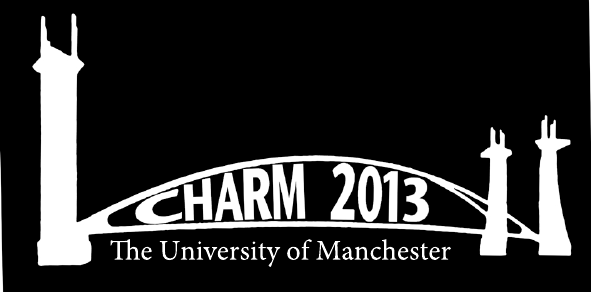}\hfill{\begin{tabular}{l} \pubnumber\\
         \pubdate  \end{tabular}}}
\newenvironment{Abstract}{\begin{quotation}  }{\end{quotation}}
\newenvironment{Presented}{\begin{quotation} \begin{center} 
             PRESENTED AT\end{center}\bigskip 
      \begin{center}\begin{large}}{\end{large}\end{center} \end{quotation}}
\def\Acknowledgements{\bigskip  \bigskip \begin{center} \begin{large}
             \bf ACKNOWLEDGEMENTS \end{large}\end{center}}
\def\venue{The 6$^{th}$ International Workshop on Charm Physics\\
(CHARM 2013)\\
Manchester, UK,  31 August -- 4 September, 2013}

\def\beq{\begin{equation}}
\def\eeq#1{\label{#1}\end{equation}}
\def\eeqn{\end{equation}}

\def\beqa{\begin{eqnarray}}
\def\eeqa#1{\label{#1}\end{eqnarray}}
\def\eeqan{\end{eqnarray}}

\let\bar=\overbar

\def\Dslash{\not{\hbox{\kern-4pt $D$}}}
\def\dslash{\not{\hbox{\kern-2pt $\del$}}}

\def\msb{{\bar{\ssstyle M \kern -1pt S}}}

\begin{document}
\begin{titlepage}
\pubblock

\vfill
\Title{\title}
\vfill
\Author{\speaker\SupportedBy{\support}\OnBehalf{\onbehalfof}}
\Address{\affiliation}
\vfill
\begin{Abstract}
We study the masses of the low-lying charmed mesons within the framework of heavy-hadron chiral perturbation theory. We work to third order in the chiral expansion,
 where meson loops contribute. In contrast to previous approaches, we use physical meson masses in evaluating these loops. This ensures that their imaginary parts are
 consistent with the observed widths of the D-mesons. The lowest odd- and even-parity, strange and nonstrange mesons provide enough constraints to determine only 
certain linear combinations of the low-energy constants in the effective Lagrangian. We comment on how lattice QCD could provide further information to disentangle
 these constants.\end{Abstract}
\vfill
\begin{Presented}
\venue
\end{Presented}
\vfill
\end{titlepage}
\def\thefootnote{\fnsymbol{footnote}}
\setcounter{footnote}{0}
%

\section{Introduction}
The masses and widths of the charmed mesons in the odd and even parity sectors have been experimentally determined, for summaries, see Refs.~\cite{pdg12,cdgn12}.
These patterns and interactions of the charmed mesons are governed by the spin symmetry of the heavy quark and the chiral symmetry $SU(3)_L\times SU(3)_R$ of the light quarks. 
Incorporating both approximate symmetries in a single framework was achieved by defining the heavy hadron chiral perturbation theory (HH$\chi$PT) \cite{ms05}.
This effective theory can be used to study dynamics of mesons containing a single heavy quark. 

Using this theory, the contributions to the physical masses of odd and even parity  $D$-mesons up to the one-loop chiral corrections were calculated by Mehen and Springer \cite{ms05} and  Ananthanarayan \textit{et al.}~\cite{absu07}.
The effective chiral Lagrangian contains more unknown low energy constants (LECs)  than the number of  experimentally known meson masses.
Thus getting unique numerical values of the coefficients is impossible. 
Mehen and Springer \cite{ms05} and Ananthanarayan \textit{et al.}~\cite{absu07} fitted expressions that depend nonlinearly on these constants and found multiple solutions, often with quite different numerical values for them. As a result, no clear pattern emerged from these fits.

In this paper, we attempt  to remove this ambiguity by following a different approach to fit these parameters.
We use the physical values of the masses  in evaluating the chiral loops. 
As a consequence, the energy of any unstable particle is placed correctly relative to the decay threshold and the imaginary part of the loop integral can be related to the experimental decay width.
The second effect is to reduce the number of unknown parameres in comparison with the current experimental data on meson masses. 
Masses at tree level depend only on certain linear  combinations of LECs. By using physical masses in chiral loops, the masses still depend linearly on these combinations.
Therefore, one can express these parameters directly in terms of the physical masses and loop integrals. 

The numerical values generated in this manner include contribution from orders beyond
$O(Q^3)$. These include divergences that
we cannot cancel using counterterms in our Lagrangian. We therefore choose to use their $\beta$-functions to estimate the large size emerging from higher-order terms.

Finally, we have used corrected expressions for the chiral loop functions, as given in Bernard \index{et al.} \cite{bkm95} and Scherer \cite{sch05}, in contrast to  the expressions presented in \cite{ms05,cas97} which are missing some finite pieces. 
\newpage
\section{Masses  of the charmed mesons within HH$\chi$PT}
We begin by writing the most general expression of the heavy-hadron chiral Lagrangian up to the order $O(Q^3)$ \cite{ms05}:
\begin{eqnarray}\label{M3}\nonumber
{\bf {\mathcal{L}}}&=&-\mathrm{Tr}[\overline{\mathcal{H}_a}\left(i v\cdot D_{ba} -\delta_H \delta_{ab}\right) \mathcal{H}_b]+\mathrm{Tr}[\overline{\mathcal{S}}_a\left(i v\cdot D_{ba} -\delta_S \delta_{ab}\right) \mathcal{S}_b]\\\nonumber
&&+ g \mathrm{Tr}[\overline{\mathcal{H}}_a\mathcal{H}_b\not{u_{ba}} \gamma_5]+g^{\prime}  \mathrm{Tr}[\overline{\mathcal{S}}_a\mathcal{S}_b\not{u_{ba}}\gamma_5]+ h \mathrm{Tr}[\overline{\mathcal{H}}_a\mathcal{S}_b\not{u_{ba}}\gamma_5 + h.c.]\\\nonumber
&&-\frac{\Delta_H}{8}\mathrm{Tr}[\overline{\mathcal{H}}_a\sigma^{\mu \nu}\mathcal{H}_a\sigma_{\mu\nu}]+\frac{\Delta_S}{8}\mathrm{Tr}[\overline{\mathcal{S}}_a\sigma^{\mu \nu}\mathcal{S}_a\sigma_{\mu\nu}]\\\nonumber
&&+a_H \mathrm{Tr}[\overline{\mathcal{H}}_a\mathcal{H}_b] m^{\xi}_{ba}-a_S \mathrm{Tr}[\overline{\mathcal{S}}_a\mathcal{S}_b] m^{\xi}_{ba} + \sigma_H \mathrm{Tr}[\overline{\mathcal{H}}_a\mathcal{H}_a] m^{\xi}_{bb}-\sigma_S \mathrm{Tr}[\overline{\mathcal{S}}_a\mathcal{S}_a] m^{\xi}_{bb}\\\nonumber
&&-\frac{\Delta^{(a)}_H}{8}\mathrm{Tr}[\overline{\mathcal{H}}_a\sigma^{\mu\nu}\mathcal{H}_b \sigma_{\mu\nu}]m^{\xi}_{ba}+\frac{\Delta^{(a)}_S}{8}\mathrm{Tr}[\overline{\mathcal{S}}_a\sigma^{\mu\nu}\mathcal{S}_b\sigma_{\mu\nu}]m^{\xi}_{ba}\\\nonumber
&&-\frac{\Delta^{(\sigma)}_H}{8}\mathrm{Tr}[\overline{\mathcal{H}}_a\sigma^{\mu\nu}\mathcal{H}_a\sigma_{\mu\nu}]m^{\xi}_{bb}+\frac{\Delta^{(\sigma)}_S}{8}\mathrm{Tr}[\overline{\mathcal{S}}_a\sigma^{\mu\nu}\mathcal{S}_a\sigma_{\mu\nu}]m^{\xi}_{bb},\nonumber
\end{eqnarray}
where $\mathcal{H}_a$ and $\mathcal{S}_a$ are the heavy meson field multiplets of the ground-state doublets and lowest lying excited-state doublets respectively. 
The members of the ground-state doublets are pseudoscalar mesons $J^P=0^-$ ($D^0,D^+,D^+_s$) and  vector mesons $J^P=1^-$ ($D^{*0},D^{*+},D^{*+}_s$), and  
the members of the excited-state doublets are  scalar meson $J^P=0^+$ ($D^0_0,D^+_0,D^+_{0s}$) and axial vector mesons $J^P=1^+$ ($D^{0'}_1,D^{1'}_1,D^{0'}_{1s}$). The index $a$ 
denotes the flavor of the light quark.
The hyperfine splitting ($\Delta$) breaks heavy-quark symmetry and is counted as $O(m_\pi)$.
The residual masses $\delta_D$ and $\delta_S$ respect both symmetries and are counted as $O(m_\pi)$. 
The terms in the second line represent the axial part of the heavy hadron chiral Lagrangian. They describe the axial vector coupling of the heavy meson fields to light mesons $\pi$, $K$, $\eta$ 
that are contained in the axial field $u_{\mu}$.

\begin{figure}[h]
\centering
\includegraphics[scale=0.5]{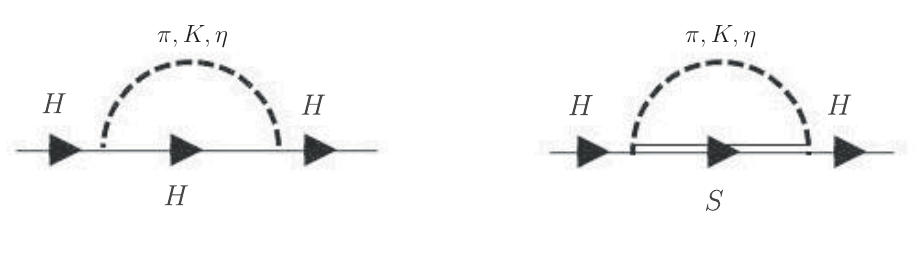}
\caption{The self-energy diagrams for the ground-state fields $H$.}
\label{fig1}
\end{figure}

\begin{figure}[h]
\centering
\includegraphics[scale=0.63]{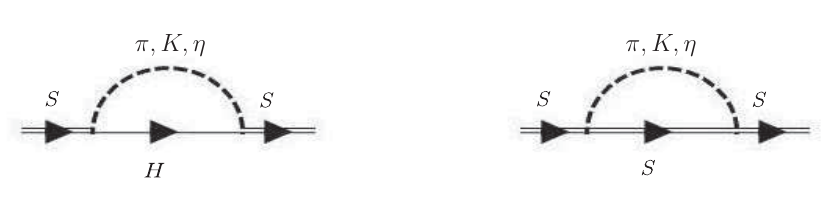}
\caption{The self-energy diagrams for the excited-state fields $S$.}
\label{fig2}
\end{figure}

From the chiral Lagrangian, one can derive the mass formula for positive and negative parity charmed mesons up to one loop self-energy $\Sigma_{A^J_a}$,
\begin{eqnarray}\nonumber
m^r_{A^J_a}= \delta_A+a_A m_a+\sigma_A \overline{m}+\frac{c_J}{4}(\Delta_A+\Delta^{(a)}_A m_a+\Delta^{(\sigma)}_A \overline{m})+\Sigma_{A^J_a},
\end{eqnarray}
where the label $A=H,S$ stand for odd and even parity states respectively, and $J$ is the total angular
momentum of the state $A$. The factors in the forth term are $c_1=1$, $c_0=-3$. The one-loop contributions to the self-energy $\Sigma_{A^J_a}$ are of order $O(Q^3)$. The Feynman diagrams of the one-loop correction to the masses of $D$ mesons are shown in the Fig.~\ref{fig1} and Fig.~\ref{fig2}.
\section{Linear combinations of the low energy constants}
The chiral Lagrangian  has a large number of unkown LECs in comparison with available experimental data.
Thus finding  unique values for them from experiments is impossible. The best we can do is to reduce their number  by grouping them into linear combinations that can be determined uniquely. 
This  procedure  is based on  the syemmtry patterns. For example, 
the  linear combinations that respect flavor symmetry are \cite{mhmb}
\begin{eqnarray}\nonumber
 \eta_A&=&\delta_A+(\frac{a_A}{3}+\sigma_A) \overline{m},\\\nonumber
\xi_A&=&\Delta_A+(\frac{\Delta^{(a)}_A}{3}+\Delta^{(\sigma)}_A) \overline{m},
\end{eqnarray}
where $\delta_A, \Delta_A$ respect chiral symmetry, but the other terms contain the average of the quark masses $\overline{m}$ which breaks it. 
The terms left after constructing  $\eta_A$ and $\xi_A$ are 
\begin{eqnarray}\nonumber
L_{A_f}=(m_s-m_n)a_A,
\end{eqnarray}
\begin{eqnarray}\nonumber
L_{A_{fs}}=(m_s-m_n)\Delta^{(a)}_A,
\end{eqnarray}
The combinations $L_{A_f}$ and $L_{A_{fs}}$ break flavor symmetry, and the latter breaks  spin symmetry.
In terms of  linear combinations, the masses can be written as \cite{mhmb}
\begin{eqnarray}\nonumber
m^r_{A^J_a}=\eta_A+\frac{c_J}{4}\xi_A+c_a L_{A_f}+c_a \frac{c_J}{4}L_{A_{fs}}+\Sigma_{A^J_a},
\end{eqnarray}
where the factor $c_a$ is $c_n=-\frac{1}{3}$ for nonstrange and  $c_s=\frac{2}{3}$ for strange mesons. Now the number of parameters,  $\xi_{H;S}, \eta_{H;S}, L_{{H;S}_f}, L_{{H;S}_{fs}}$, is eight which is
 equal to the number of observed low-lying $D$-meson states.
\section{Results and Discussion}
In our fitting, the one-loop self energy $\Sigma_{A^J_a}$ is evaluated with physical values of the charmed meson masses.
This gives the energy of the decaying particle relative to the threshold in the right place.  This ensures that the imaginary parts of the loop functions are correctly related to the experimental decay widths.
The resulting masses depend linearly on certain combinations of parameters. However, the resulting values for these parameters
 contain contributions beyond $O(Q^3)$. We have therefore used their $\beta$-functions  to estimate a theoretical error coming from higher-order terms.

The calculations are performed at the physical values of pion decay constant $f_{\pi}=92.4\mathrm{MeV}$, and of the coupling constants $g$ and $h$ that are extracted from the strong decay widths $g=0.64\pm0.075$ and $h=0.56\pm0.04$, for details see \cite{cdgn12}.
Masses of $D$-mesons are taken from the $\mathrm{PDG}$ \cite{pdg12}.
The renormalization scale $\mu$ is chosen to be the average of the pion and  kaon masses $\mu=317\mathrm{MeV}$.
The resulting numerical values of the parameters inhabit the odd parity sector with associated uncertainties are \cite{mhmb}
\[\eta_H=171.614\pm44\pm5 \, \mathrm{MeV},\]
\[\xi_H=150.946\pm5\pm6 \,\mathrm{MeV},\]
\[ L_{H_f}=242.904\pm40\pm18\,\mathrm{MeV},\]
\[L_{H_{fs}}=-52.3251\pm19\pm15\,\mathrm{MeV},\]
where the first uncertainty is  the experimental error associated with physical masses of charmed mesons, and the
second uncertainty is the theoretical error that we have estimated from the $\beta$-functions.

\begin{figure}[h]
\centering
\includegraphics[scale=0.8]{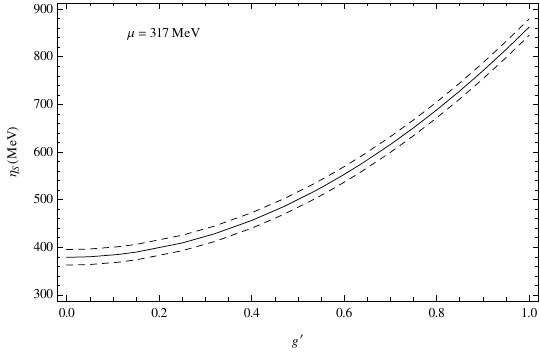}
\caption{The combination $\eta_S$ is plotted against $g^{\prime}$. The main value is represented by the solid line.
The experimental errors are shown by the two-dashed lines. The theoretical uncertainty is a constant  $\pm 5\mathrm{MeV}$.}\label{fig3}
\end{figure}

\begin{figure}[h]
\centering
\includegraphics[scale=0.8]{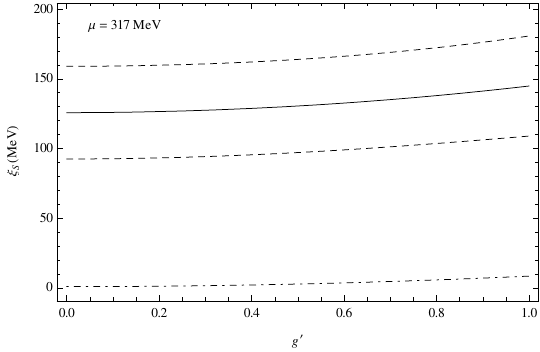}
\caption{
The combination $\xi_S$ is plotted against $g^{\prime}$.
The experimental uncertainties are shown by two dashed lines surrounding the main values and the theoretical
uncertainty is shown by dot-dashed line.}\label{fig4}
\end{figure}

\begin{figure}[h]
\centering
\includegraphics[scale=0.8]{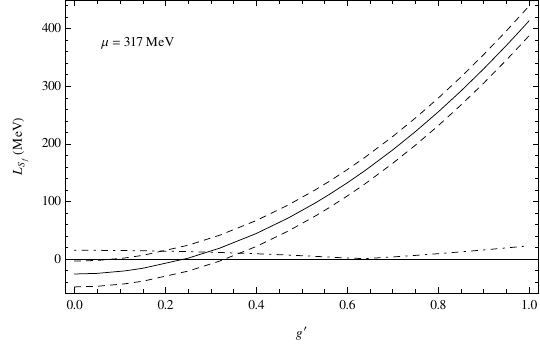}
\caption{The combination $L_{S_f}$ is plotted against $g^{\prime}$. The experimental and the theoretical errors are shown by dashed, and dot-dashed lines respectively.}\label{fig5}
\end{figure}

\begin{figure}[h]
\centering
\includegraphics[scale=0.8]{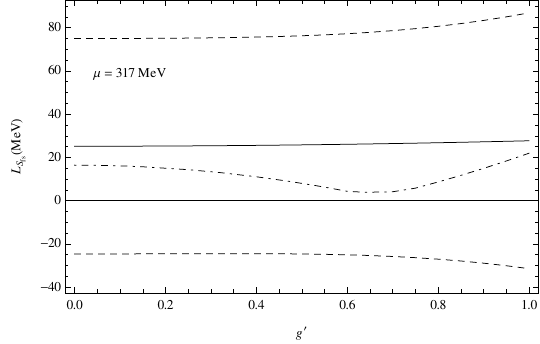}
\caption{The combination $L_{S_{fs}}$ is plotted against $g^{\prime}$. The experimental and the theoretical errors are shown by dashed, and dot-dashed lines respectively.}\label{fig6}
\end{figure}

The situation for the even-parity parameters is different because the coupling constant $g^{\prime}$ is not determined.
Since the value of the odd parity coupling constant is $0.64$, it is plausible to consider values for  $g^{\prime}$ in the range $0$ to $1$.
The correlations between $g^{\prime}$ and $\eta_S,\xi_S,L_{S_f},L_{S_{fs}}$ are shown in Figures~\ref{fig3}-~\ref{fig6}\, \cite{mhmb}. The plots also show the associated experimental and theoretical errors.

Experimental information is not sufficient to separate the combinations of parameters into pieces that respect and break chiral symmetry, which limits their usefulness for applications 
to other observables. Lattice  $\mathrm{QCD}$ calculations would be required to perform further separations of terms.
For example, to disentangle chirally symmetric coefficients $\delta_H, \Delta_H$  from chirally breaking terms, lattice calculations with different quark masses are needed for charmed mesons ground-state and excited-state.

\Acknowledgements
I am grateful to my supervisor, Prof. Mike Birse, whose guidance and support helped  me to develop an understanding of the subject.

\end{document}